\documentclass[aps,pre,twocolumn,showpacs,floatfix,superscriptaddress]{revtex4}
\usepackage{epsfig,amssymb,amsmath,amsthm,latexsym,rotating,setspace,graphicx,float}

\usepackage{color}
\usepackage[utf8]{inputenc}

\begin{document} 

\title{Statistical theory of quasi stationary states beyond the single water-bag case study} 

\author{Mallbor Assllani}
\email{malborasllani@yahoo.com}
\affiliation{Dipartimento di Energetica, Universit\'a di Firenze, via Santa Marta, 3 - I-50139 Firenze, Italy}

\author{Duccio Fanelli}
\email{duccio.fanelli@gmail.com}
\affiliation{Dipartimento di Energetica, Universit\'a di Firenze, INFN and CNISM via Santa Marta, 3 - I-50139 Firenze, Italy}
\affiliation{Centro Interdipartimentale per lo Studio delle Dinamiche Complesse, via Sansone, 1 - I-50019 Sesto Fiorentino, Italy}

\author{Alessio Turchi}
\email{alessio.turchi@gmail.com}
\affiliation{Dipartimento di Sistemi e Informatica, Universit\'a di Firenze,via Santa Marta, 3 - I-50139 Firenze, Italy}
\affiliation{Centre de Physique th\'eorique, Aix-Marseille Universit\'e,  Campus de Luminy, F-13288 Marseille Cedex 09, France}
\affiliation{Centro Interdipartimentale per lo Studio delle Dinamiche Complesse, via Sansone, 1 - I-50019 Sesto Fiorentino, Italy}

\author{Timoteo Carletti}
\email{timoteo.carletti@fundp.ac.be}
\affiliation{naXys, Namur Center for Complex Systems and University of Namur,  Namur, Belgium 5000}

\author{Xavier Leoncini}
\email{Xavier.Leoncini@cpt.univ-mrs.fr}
\affiliation{Centre de Physique th\'eorique, Aix-Marseille Universit\'e,  Campus de Luminy, F-13288 Marseille Cedex 09, France}

\begin{abstract}
An analytical solution for the out-of-equilibrium quasi-stationary states of the paradigmatic Hamiltonian Mean Field (HMF) model can be obtained from a maximum entropy principle. The theory has  been so far tested with reference  to a specific class of initial condition, the so called ({\it single-level}) water-bag type. In this paper a step forward is taken by considering an arbitrary number of overlapping water bags. The theory is benchmarked to direct microcanonical simulations performed for the case of a {\it two-levels} water-bag. The comparison is shown to return an excellent agreement. 
\end{abstract} 


\maketitle

\section{Introduction}
\label{uno}

Long range interacting systems (LRS) are becoming a popular topic of investigation \cite{review} in physics, due to the rich and intriguing phenomenology that they display. A system is said to fall in the realm of LRS if the two body potential scales as $r^{-\alpha}$ with $\alpha < d$, and where $r$ stands for the inter-particle distance and $d$ the dimension of the embedding space. Several physical systems share this property, which ideally embraces distinct domains of applications. Gravity \cite{galaxy} is certainly the most spectacular 
example among the wide gallery of systems governed by long range interactions, but equally important are the cases of 
turbulence \cite{vortex}, plasmas \cite{plasma} and wave-particle interactions \cite{spinwave,fel}. 

Peculiar and counterintuitive thermodynamics features manifest in LRS: negative specific heat can occasionally develop in the 
microcanonical ensemble, close to first order phase transitions, a surprising fact first discovered in astrophysical context, that seeds statistical ensemble inequivalence. As concerns the dynamics, LRS have been reported to experience a very slow relaxation towards the 
deputed thermodynamic equilibrium. Indeed, they can be trapped in long lasting out-of-equilibrium phases called Quasi Stationary 
States (QSSs). The lifetime of the QSSs diverges with the systems size $N$. Interestingly, it displays different scaling
 behaviors versus $N$, which range from exponential to power law, being relic of the specific initial condition selected. 
As a consequence, the orders the limits $N \rightarrow \infty$ and  $t \rightarrow \infty$ are taken do matter. Performing
 the continuum limit before the infinite time limit, implies preventing the system from eventually attaining its equilibrium 
and so freezing it indefinitely in the QSS phase.    

In physical applications where long-range couplings are at play, as the ones mentioned above, the number of elementary
 constituents composing the system being examined, is generally large. The time of duration of the out-of-equilibrium phase
 can therefore be exceedingly long, definitely longer the the time of observation to which experimentalists are bound. Given 
this scenario, it is of paramount importance to develop dedicated analytical strategies to gain quantitative insight into
 the complex and diverse zoology of the QSSs, as revealed by direct numerical simulations. Working along these lines, it was
 shown that QSS can be successfully interpreted as equilibria of the collisionless Vlasov equation which appears to rule the
 dynamics of a broad family of long range models, when recovering the continuum picture from the governing discrete
 formulation. The average characteristics of the QSS, including the emergence of out-of-equilibrium transitions, can be
 analytically predicted via a maximum entropy variational principle, pioneered by Lynden-Bell in \cite{LB} and more recently
 revisited with reference to paradigmatic long-range applications \cite{Antoniazzi, fel, fel1}.\\
As we shall clarify in the forthcoming sections, the predictive adequacy of the Lynden-Bell violent relaxation theory 
has been so far solely assessed for a very specific class of initial conditions. These are the so called (single) water-bags:
 particles are assumed to initially populate a bound domain of phase space and therein distributed with a uniform probability.
 The aim of this paper is to take one simple step forward and challenge the validity of the theory when particles are instead
 distributed within two (uniformly filled) levels.  In principle, any smooth profile could be approximated by a piecewise
 function, made of an arbitrary number of collated water-bags \cite{Gyrowb}. Our idea is to perform a first step towards the generalized
 multi-levels setting, by first evaluating the formal complexity of the procedure involved and then drawing a direct 
comparison with the numerics relative to the two level case.  

To accomplish this task, we will focus on the celebrated Hamiltonian Mean Field model, often referred to as the representative model of long range interactions. The HMF describes the motion on a circle of an ensemble of $N$ rotors mutually coupled via an all-to-all cosines like potential. In the continuum limit, the single particle distribution function obeys to the Vlasov equation, the driving potential being self-consistently provided by the global magnetization, namely the degree of inherent bunching. QSSs exist for the HMF model and have been deeply studied, both with analytical and numerical means, for the single water-bag case.
 
The paper is organized as follows. In the next section we shall introduce the discrete HMF model and discuss its continuum  Vlasov based representation. We will also introduce the basic of the violent relaxation approach. Then, in section III, we shall turn to 
discussing the generalized water bag setting, solving, in section IV, the corresponding variational problem. We will then specialize in section V on the two levels case and compare the theory predictions to the simulations. Finally in section VI we will sum up and conclude.

\section{The HMF model}

The HMF model describes the dynamics of $N$ particles (rotors) moving on a circle and interacting via a mean field potential which is self-consistently generated by the particles themselves. Formally, the HMF is defined by the following Hamiltonian:
\begin{equation}
\label{hmfeq}
\mathcal{H} = \sum_{i=1}^N \frac{p_i^2}{2} + \frac{1}{2N}\sum_{i,j=1}^N(1- \cos(\theta_i-\theta_j)),
\end{equation}               
where $\theta_i$ identifies the position of particle $i$ on the circle and
$p_i$ is the canonically conjugated momentum. Because the interactions are not bounded to a small number of
neighboring particles, the interaction is all-to-all thus the potential
is inherently long range. \\

Starting from the water bag initial condition, the HMF system experiences a fast relaxation towards an intermediate regime, before the final equilibrium is eventually attained. This metastable phase is a Quasi Stationary State (QSS), the out-of-equilibrium transient to which we have alluded to in the introduction. The lifetime of the QSS is shown to diverge with the system size $N$, an observation that has non-trivial consequences when one wishes to inspect the continuum ($N \to \infty$) limit. QSSs are in fact stable, attractive equilibria of the continuous analogue of the discrete Hamiltonian picture, and bear distinctive traits that make them substantially different from the corresponding equilibrium solutions. 

To monitor the dynamics of the system it is customary to record the time evolution of the magnetization. 
This latter is defined as:

\begin{equation}
\mathbf{M} =\sum_{i=1}^N e^{i\theta}.
\end{equation}

It is a complex quantity whose modulus $M$ measures the degree of bunching of the distribution of particles. Depending on the selected characteristics of the initial (single) water-bag, the system can evolve towards an (almost) homogeneous QSS or, conversely, result in a magnetized phase. The swap between the two regimes can be understood as a genuine phase transition, with energy and initial magnetization playing the role of control parameters. 

It can be rigorously shown \cite{HMFV} that, in the continuum limit, the HMF system is formally described by the Vlasov equation, which governs the evolution of the single particle distribution function $f(\theta,p,t)$. In formula:
\begin{equation}
\label{Vlasoveq}
\frac{\partial f}{\partial t} + p \frac{\partial f}{\partial \theta} - \left( M_x[f] \sin \theta - M_y[f] \cos \theta \right) 
\frac{\partial f}{\partial p} = 0,
\end{equation}
where $M_x=\int f \cos \theta d \theta d p$ and $M_y=\int f \sin \theta d \theta d p$ are the two components of the magnetization $\mathbf{M}$. With reference to cosmological applications, Lynden-Bell proposed an analytical approach to determine the stationary solutions of the Vlasov equation, pioneering the theory that it is nowadays referred to as to the violent relaxation theory. He first considered the coarse grained distribution $\bar{f}$, obtained by averaging the microscopic $f(\theta,p,t)$ over a finite grid. Then, the key idea is to associate to $\bar{f}$ a mixing entropy $S[\bar{f}]$, via a rigorous counting of the microscopic configurations that are compatible with a given macroscopic state. The steps involved in the derivation are highlighted in the remaining part of this section. 

Let us label with $\rho(\theta,p,\eta)$ the probability density of finding the level of phase density $\eta$ in the neighborhood of the position $[\theta, p]$ in phase space. We are here implicitly assuming to deal with a continuum spectrum of allowed levels. Following Lynden-Bell the coarse grained locally averaged single particle distribution reads:
\begin{equation}
\bar{f}(\theta,p)=\int \rho(\theta,p,\eta)\eta \, d\eta\:.
\end{equation}

The Vlasov equation, which rules the dynamics of $\bar{f}$, conserves the hypervolumes $\nu(\eta)$ associated 
to each of the selected levels \cite{Chavanis}. Mathematically, the quantities $\nu(\eta)$ reads:
\begin{equation}
\nu(\eta) = \int \rho(\theta, p, \eta) d \theta d p\:,
\end{equation}
and are therefore invariant of the dynamics. Assume now to deal with a
discrete set of $n$ distinct levels. The probability density function reads therefore:
\begin{equation}
\label{rhoNlevels}
\begin{split}
\rho_n\left(\theta, p,\eta\right)=&\sum_{i=1}^{n} \bar{f}(\theta,p) \alpha_i \delta(\eta - f_i)\ + \\
&\left(1-\sum_{i=1}^{n} \bar{f} (\theta,p) \alpha_i \right)\delta(\eta),
\end{split}
\end{equation}
where $f_i$ refers to the value of the $i$-th level and $\alpha_i$ stands for
the portion of phase space that is hosting the selected level. The
  rightmost term in the previous equation, stands thus for the background. We
shall return on these aspects later on, when commenting on the underlying
normalization condition.   

Starting from this setting, Lynden-Bell suggested to divide the phase space macrocell $[\theta,\theta + d\theta;p, p + dp]$ 
into microcells each occupied by just one of the allowed levels. The Lynden-Bell mixing entropy is obtained as the
 logarithm of the total number of microstates associated to a given macrostate. 
 A combinatorial calculation \cite{Chavanis} yields to:

\begin{equation}
\label{entrorhoeta}
\begin{split}
S = -\int\rho_n(\theta, p,\eta)\ln\rho_n(\theta, p,\eta) \, d\theta dp d\eta,
\end{split}
\end{equation}

The statistical equilibrium of the system follows from maximizing the entropy functional $S$
while imposing the constraints of the dynamics: energy, momentum and normalization are in fact conserved 
quantities, as well as the quantities $\nu(\eta)$. In the following we shall discuss a specific class of initial condition, the multi-level waterbags, which naturally extends beyond the single water bag case study, so far explicitly considered in the literature. It is our intention to test the predictive ability of the Lynden-Bell theory within such generalized framework. The theory will be developed with reference to the general setting, including $n$ levels. The benchmark with direct simulations will be instead limited to the two-levels case, i.e. $n=2$.

\section{The generalized water-bag}

The single water-bag initial condition takes a constant value $f_1$ within a 
finite portion of the phase space, and zero outside of it. Although this is
the only prescription to be accommodated for, rectangular domains are usually
chosen for practical computational reasons. Following
\cite{Antoniazzi}, we shall label $[\Delta \theta, \Delta p]$ the widths of
such a rectangle, as calculated respectively along $\theta$ and $p$
directions. A second simplification
is also customarily be assumed: the rectangle
 is centered in the origin, so that  
 $\theta \in [-\frac{\Delta \theta}{2},\frac{\Delta \theta}{2}]$ and $p \in [-\frac{\Delta p}{2},\frac{\Delta p}{2}]$.

By operating in this context, the Lynden-Bell variational problem studied in e.g. \cite{Antoniazzi} is shown to 
yield to a Fermionic stationary distribution, which successfully enables to capture some of the essential 
traits of the QSS. These includes an accurate characterization of the
out-of-equilibrium transitions from magnetized to non-magnetized QSS. First
and second order phase transitions, that merge in a tricritical point, were in
fact singled out for the HMF model, a theoretical prediction confirmed by
direct numerical inspection. As stated above, the general philosophy that
inspires the Lynden-Bell theory is however broader than the specific realm to
which it was relegated and its potentiality deserves to be further
clarified. We will here extend the treatment to the multi-levels water bag
initial condition, a step that opens up the
  perspective to eventually handle
 more realistic scenarios, where smooth
  distributions could be considered.

\begin{figure}
\includegraphics[draft=false,clip=true, width=8cm]{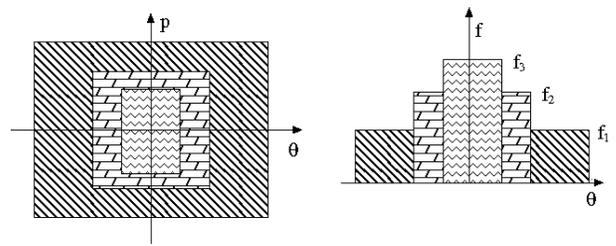}
\caption{Pictorial representation of a tree-levels ($n=3$) water-bag initial condition. }
\label{figure:0}
\end{figure}

Following the notation introduced above, the arbitrary integer $n$ quantifies the total number of distinct levels 
that are to be allowed for, when considering the generalized initial distribution function $f_{init}$. 
Arguably, by accounting for a large enough  collection of independent and discrete levels, one can approximately
 mimic any smooth profile. A pictorial representation of the family of initial conditions to which we shall
 refer to in the forthcoming sections when discussing the specific case study $n=3$ is depicted in figure \ref{figure:0}.

Mathematically,  the initial distribution function ${f}_{init}$ can be written as: 
\begin{equation}
\label{f2lvl}
{f}_{init}(\theta,p)=
\begin{cases}
f_j & \text{if $\theta \in \mathbf{\Theta_j}$ and $p \in \mathbf{P_j}$},\\
f_0 = 0 & \text{elsewhere},
\end{cases}
\end{equation}
Here $\Gamma_j=[\Theta_j, P_j]$, $j=1,\dots, n$ identifies the  domain in phase space associated to level $f_j$. The corresponding volume is labeled $\alpha_j$. 

A simple algebraic manipulation starting from the definition
  of the probability density (\ref{rhoNlevels}) yields to:  
\begin{equation}
\label{normcond}
\int f_{init}(\theta,p)\, d\theta dp = \int\rho_n(\theta,p,\eta) \eta \, d\theta dp d\eta = \sum_{j=0}^nf_j \alpha_j = 1.
\end{equation}
The scalar relation (\ref{normcond}) links together the $2n$ constants, $f_j$ and $\alpha_j$, that are to 
be assigned to fully specify the initial condition. In other words, only 
$2n-1$ scalars are needed to completely parameterize the initial condition.  
Importantly, the single water bag limit is readily recovered once the phase
space support of the levels indexed with $j$ other than $j=1$ shrinks and eventually fades out. This condition implies requiring $\alpha_j \rightarrow 0$ 
  for $j>1$. In formula:
\begin{equation}
\label{limitrho}
\rho_1=\lim_{\alpha_j\to 0,j>1}\rho_n =\bar{f} \alpha_1 \delta\left(\eta - f_1\right) + \left(1-\bar{f} \alpha_1\right)\delta\left(\eta\right).
\end{equation}
Moreover, by making use of the normalization condition one gets
$\alpha_1=\frac{1}{f_1}$, which, inserted (\ref{limitrho}), returns
immediately the well known form of the one level density distribution  
function. Notice that the initial value of the macroscopic observables can 
be computed from the explicit knowledge of $\rho_1$. 

\section{The generalized n-levels equilibrium}
\label{tre}

The QSS distribution function $\bar{f}_{eq}(\theta,p)$  for the HMF model, relative to the generalized $n$-levels water-bag
 initial condition, is found by maximizing the Lynden Bell entropy, under the constrains of the dynamics. This in turn
 implies solving a variational problem. The solution is relative to the microcanonical ensemble, since 
the Vlasov equation implies that  we work with fixed total energy.\\

From equation (\ref{entrorhoeta}) and (\ref{rhoNlevels}), the generic $n$-levels entropy takes the following functional form:
\begin{equation}
\label{entro2lvl}
\begin{split}
S[\bar{f}] = & - \int \lbrace\sum_{j=1}^n \bar{f}\alpha_j\ln(\bar{f}\alpha_j) + \\
 &+ (1-\sum_{j=1}^n \bar{f}\alpha_j )\ln(1-\sum_{j=1}^n \bar{f}\alpha_j )\rbrace \, d\theta dp
\end{split}
\end{equation}
The conserved quantities are respectively the energy $E$: 
\begin{equation}
\label{encon}
E\left[ \bar{f}\right]=\int\frac{p^2}{2} \bar{f}(\theta,p) \, d\theta dp -\frac{M[\bar{f}]^2 -1}{2} \equiv E_{n}\, ,
\end{equation}
and the total momentum $P$
\begin{equation}
\label{momcon}
P[ \bar{f} ]=\int \bar{f}(\theta,p) p \, d\theta dp \equiv P_{n}\,.
\end{equation}
The scalar quantity $E_n$ relates to the geometric characteristics of the bounded domains that define our initial condition. Conversely, as we will be dealing with patches $\Gamma_j$ symmetric with respect to the origin, 
one can immediately realize that $P_n=0$. \\ 

The $n$ volumes of phase space, each deputed to hosting one of the considered levels, are also invariant of the dynamics. We have therefore to account for the conservation of $n$ additional quantities, the volumes $\nu_j[ \bar{f}]$ for $j=1,..n$, defined as: 
\begin{equation}
\label{masscon}
\nu_j[ \bar{f}]=\int \bar{f}(\theta,p)\alpha_j \, d\theta
dp\, ,
\end{equation}

Moreover using the normalization
condition for the coarse grained distribution function
$\bar{f}(\theta,p)$, we get $\nu_j[ \bar{f}]=\alpha_j$. Equivalently,
by imposing the above constraints on the hypervolumes, we also guarantee the
normalization of the distribution function, which physically amounts to impose
the conservation of the mass.  

Summing up, the variational problem that needs to be solved to eventually recover the stationary distribution $\bar{f}_{eq}(\theta,p)$ reads: 
\begin{equation}
\label{varprob}
\max_{\bar{f}} \lbrace S[\bar{f}]\; |\; E\left[\bar{f}\right]=E_n; P\left[\bar{f}\right]=P_n; \nu_i\left[\bar{f}\right]=\alpha_i\rbrace,
\end{equation}
where the entropy functional $S[\bar{f}]$ is given by eq. (\ref{entro2lvl}). This immediately translates into:
\begin{equation}
\label{lagrange}
\delta S - \beta\delta E - \lambda\delta P - \sum_{j=1}^n\mu_j\delta\nu_j=0.
\end{equation}
where $\beta$, $\lambda$ and $\mu_j$ stands for the Lagrange multipliers associated respectively to energy, momentum and volumes (or equivalently mass) conservations. \\

A straightforward calculation yields to the following expression for  $\bar{f}_{eq}(\theta,p)$:

\begin{equation}
\label{fbar}
\bar{f}_{eq}=\frac{1}{B+Ae^{\beta^\prime\left(\frac{p^2}{2}-\mathbf{M}[\bar{f}_{eq}]\cdot \mathbf{m}\right)+\lambda^\prime p+\mu^\prime}}
\end{equation}
where
\begin{equation}
\label{AB}
B=\sum_{j=1}^n\alpha_j\;;\; A=\left(\prod_{j=1}^n\alpha_j^{\alpha_j}\right)^{\frac{1}{B}}
\end{equation}

where

\begin{equation}
\label{lagrangedef}
\begin{split}
\beta^\prime =& \frac{\beta}{B}, \\
\quad \lambda^\prime =& \frac{\lambda}{B}, \\
\mu^\prime =& \frac{\sum_{j=1}^n\mu_j}{B},
\end{split}
\end{equation}

and $\mathbf{m}=[\cos(\theta),\sin(\theta)]$. 

The above solution is clearly consistent with that obtained for the single
water bag case study \cite{Antoniazzi}. This latter is in fact recovered in
the limit ($\alpha_{j} \to 0$ for $j>1$ while $\alpha_1=\frac{1}{f_1}$):
\begin{equation}
\label{fbarlimit}
\lim_{\alpha_j\to 0,j>1}\bar{f}_{eq} =\frac{f_1}{1+e^{f_1\left[\beta\left(\frac{p^2}{2}-\mathbf{M}\cdot \mathbf{m}\right)+\lambda p+\mu_1\right]}}
\end{equation}
\\
Notice that the equilibrium distribution $\bar{f}_{eq}$ depends on $\mathbf{M}$, which is in turn function of  
$\bar{f}_{eq}$ itself. The two components of the magnetization, respectively $M_x$ and $M_y$ are therefore unknowns of the problem, implicitly dependent on  $\bar{f}_{eq}$. This latter is parameterized in terms of the Lagrange multipliers. Their values need  to be self-consistently singled out. As a first simplification, we observe that the specific symmetry of the selected initial condition ($P_n=0$) implies $\lambda=0$. Hence, just the two residual Lagrange multipliers are to be computed: the Lynden-Bell inverse temperature $\beta$ and the cumulative chemical potential $\mu^\prime$ \footnote{Being only interested in $\mu^\prime$ (to solve for $\bar{f}_{eq}$) 
and not on the complete collection of $\mu_j$, we can hereafter focus just on 
the conservation of the global mass, i.e. the normalization.}.
The number of unknowns total therefore to four ($M_x$, $M_y$, $\beta$, $\mu^\prime$)  and enter the following system of implicit equations for the constraints: 

\begin{align}
\label{finaleq}
&E = \frac{\tilde{A}}{2\beta^{\prime 3/2}}\int e^{\beta^\prime \mathbf{M}\cdot\mathbf{m}}F_2(y)\, d\theta - \frac{M^2-1}{2}
\\
\label{finaleq1}
&1 = \frac{\tilde{A}}{\sqrt{\beta^\prime}}\int e^{\beta^\prime \mathbf{M}\cdot \mathbf{m}}F_0(y)\, d\theta
\\
\label{finaleq2}
&M_x = \frac{\tilde{A}}{\sqrt{\beta^\prime}}\int e^{\beta^\prime \mathbf{M}\cdot\mathbf{m}}F_0(y)\cos(\theta)\, d\theta
\\
\label{finaleq3}
&M_y = \frac{\tilde{A}}{\sqrt{\beta^\prime}}\int e^{\beta^\prime \mathbf{M}\cdot\mathbf{m}}F_0(y)\sin(\theta)\, d\theta
\end{align}
Here we have expressed the relations as function of the Fermi integrals 
$F_h(y)=\int \frac{p^he^{-p^2/2}}{1+ye^{-p^2/2}}\, dp$, with 
$y=\tilde{A}Be^{\beta' \mathbf{M}\cdot\mathbf{m}}$ and $\tilde{A}=A^{-1}e^{-\mu^\prime}$.
The system of equations (\ref{finaleq}),(\ref{finaleq1}),(\ref{finaleq2}),(\ref{finaleq3}) 
can be solved numerically. In doing so one obtains a numerical value for the involved Lagrange multipliers, as well as for the magnetization components, by varying the parameters that encode for the initial condition. We numerically checked (data not shown) that in the limit of a single water bag $\alpha_{j>1} \to 0$ the solution reported in \cite{Antoniazzi} is indeed recovered. In the following section we turn to discussing the theory predictions with reference to the simple case of two water bag ($n=2$), validating the results versus direct numerical simulations. \\

\section{The case $n=2$: theory predictions and numerical simulations.}
\label{cinq}

 We  here consider the simplifying setting where two levels ($n=2$) water-bag are allowed for. We are in
 particular interested in monitoring the dependence of $M=\sqrt{M_x^2+M_y^2}$ versus the various  parameters that
 characterize the initial condition. We recall in fact that, for the case of a single 
water-bag, out of equilibrium transitions have been found
\cite{Antoniazzi}, which separates between homogeneous and magnetized
phases. A natural question is thus to understand what is
going to happen if one additional level is introduced in the initial
condition. The level $f_1$ is associated to a rectangular domain $\Gamma_1$ of respective widths 
$\Delta \theta_1$ and $\Delta p_1$. The level $f_2$ insists instead on an adjacent domain $\Gamma_2$, whose external 
perimeter is delimited by a rectangle of dimensions $\Delta \theta_2$ and $\Delta p_2$. The corresponding 
surface totals hence $\Delta \theta_2 \Delta p_2- \Delta \theta_1 \Delta p_1$. 

Recall that the energy $E_2$ ($E_n$ for $n=2$) can be estimated as dictated by formula (\ref{encon}) and reads in this specific case:

\begin{eqnarray}
\label{en2lvl}
E_2 &=& \frac{1}{24} \left( f_1 \Delta \theta_1 \Delta p_1^3 + (f_2-f_1) f_2 \Delta \theta_2 \Delta p_2^3 \right)  \\
&+& \frac{1-16(f_1 \Delta p_1 \sin \Delta \theta_1 /2 +  (f_2-f_1)\Delta p_2 \sin \Delta \theta_2/2 ) }{2}. \nonumber
\end{eqnarray}

The one-level limit is readily recovered by 
simultaneously imposing $\Delta \theta_2 \rightarrow 0$ and $\Delta
p_2\rightarrow 0$ (which also implies $\alpha_2 \rightarrow 0$). By
invoking the  normalization condition (\ref{normcond}) the following
relation holds:  
 \begin{equation}
\label{conditionA}
\lim_{\Delta \theta_2, \Delta p_2 \rightarrow 0} E_2 = \frac{1}{6}\Delta p_1^2 + \frac{1}{2}(1 - M_0^2) 
\end{equation} 
where $M_0=2 \sin (\Delta \theta_1 / 2)/\Delta \theta_1$.  The above relation coincides with the canonical
expression for $E_1$, as e.g. derived in \cite{hmforig}.

Relation (\ref{en2lvl}) enables us to estimate the energy associated to the selected initial condition and can be used in the self-consistency equations (\ref{finaleq}). Before turning to illustrate the predicted solution, 
let us note that the normalization (\ref{normcond}) reduces for $n=2$ to:

\begin{equation}
\label{norm1}
\alpha_1 f_1 + \alpha_2 f_2 = 1.
\end{equation}
 
To explore the parameter space we have decided to monitor the dependence of $M$ on $f_1$, which therefore acts as a control parameter. To this end, we proceed by fixing the quantity $\Delta f \equiv f_2-f_1$, the difference in hight of the considered levels. Furthermore, we specify the quantity $\alpha_1$, while $\alpha_2$ is calculated so to match the normalization constraint. 

\begin{figure}
\includegraphics[draft=false,clip=true, width=8cm]{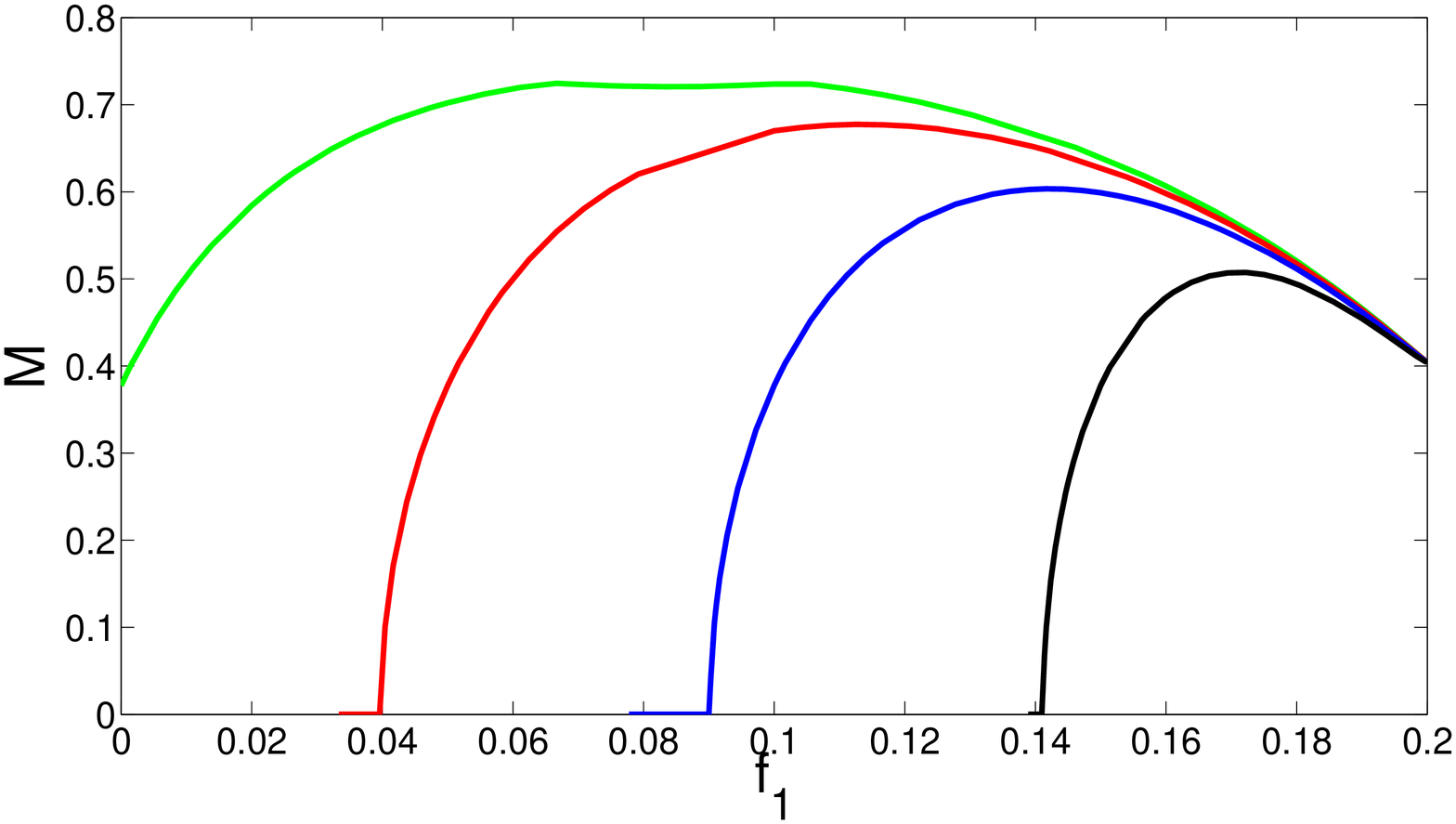}
\caption{Analytical predictions for the equilibrium magnetization $M[\bar{f}_{eq}]$ as obtained 
for different values of the initial two levels water-bag distribution. The two levels are respectively labeled 
$f_1$ and $f_2$. We here work at constant $\alpha_1=5$ and $\Delta f= f_2-f_1$, while moving the control parameter $f_1$.
 The analysis is repeated for distinct values of $\Delta f$ (from left to right $\Delta f = 0.2,0.15,0.1,0.05$ ). 
$\alpha_2$ is computed according to eq. (\ref{normcond}). For $f_1 \rightarrow 1/\alpha_1=0.2$ the normalization condition yields to
 $\alpha_2 \rightarrow 0$, and the distribution collapses to the limiting case of a single water-bag.}
\label{figure:1}
\end{figure}

The analysis is then repeated for distinct choices of $\Delta f$, 
so to eventually elaborate on the importance of such crucial parameter. The results are displayed in figure 
\ref{figure:1}. The curves collapse towards a point  that corresponds to the limiting condition $\alpha_2 \rightarrow 0$ ($f_1 = 1/\alpha_1$): this special solution is
met when the hypervolume populated by the level $f_2$  shrinks to zero, so driving the system towards 
the standard one level setting. By progressively reducing $f_1$ the predicted magnetization first increases and subsequently decreases to eventually reach zero at a critical threshold $f_1^c$. For $ f_1> f_1^c$  the system is predicted to evolve towards a magnetized, hence non homogeneous phase. Alternatively, for $0 < f_1 < f_1^c$ a homogeneous phase is expected to occur. Interestingly, the transition point
$f_1^c$ depends on the selected $\Delta f$: the larger  
$\Delta f$ the smaller the value of the transition point, corresponding to a shift to the left in figure
\ref{figure:1}. Notice that above a limiting value of $\Delta f$, which
self-consistently corresponds to imposing $\alpha_2>\Delta f$, the value of $f_1$
has to forcefully become negative so to respect the normalization condition. A
{\it well} hence opens up in phase space, an intriguing scenario that can be
formally handled within the descriptive Vlasov framework but that we have here
deliberately omitted to deepen any further. The smooth phase transition
as depicted in figure \ref{figure:1} is therefore lost above a threshold value
of $\Delta f$, when the predicted value of $M$ associated to $f_1=0$ turns out
to be greater than zero.

\begin{figure}
\includegraphics[draft=false,clip=true, width=8cm]{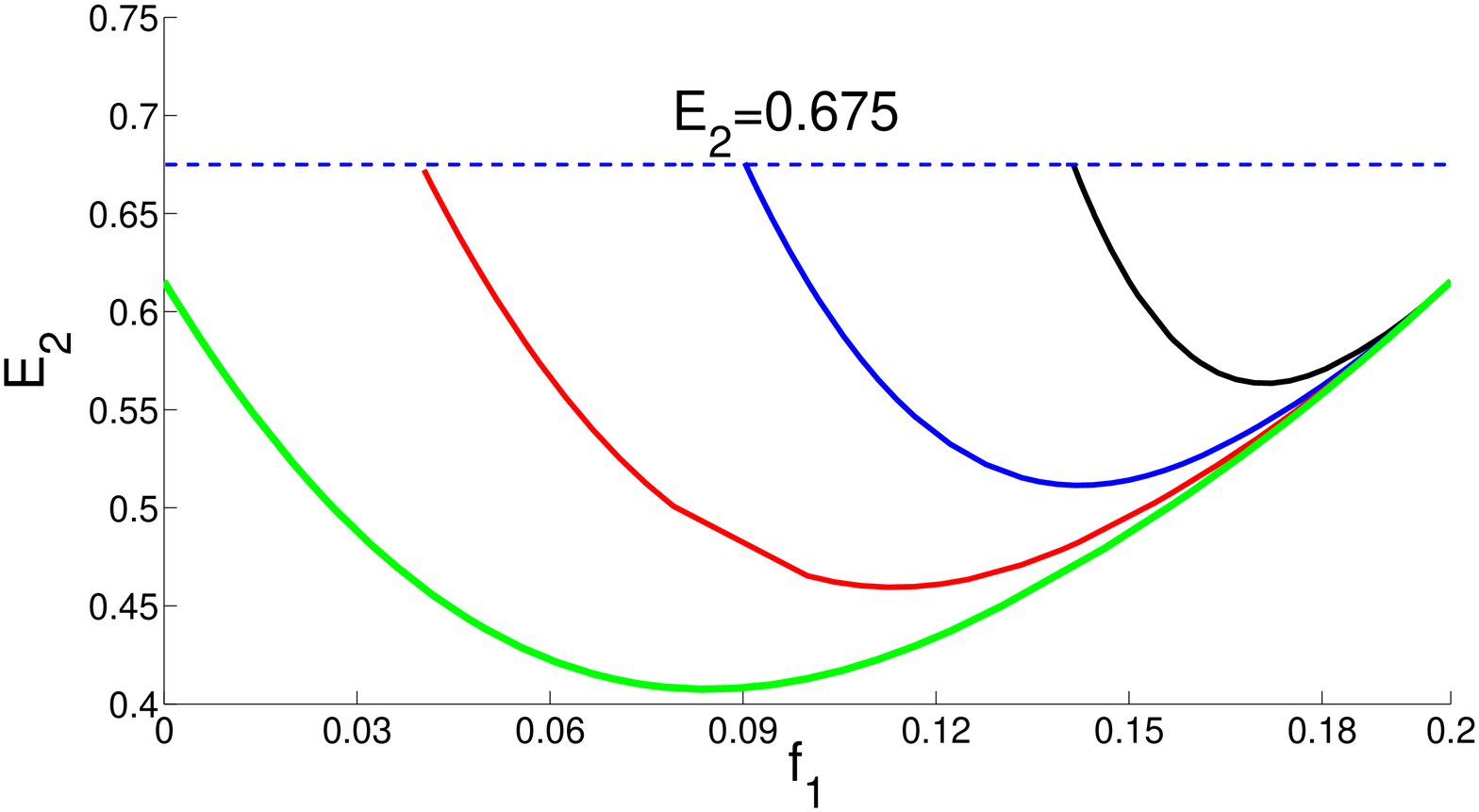}
\caption{The analytical curves (same setting as in figure (\ref{figure:1})) are now plotted in the plane ($E_2,f_1$). 
(from left to right $\Delta f = 0.2,0.15,0.1,0.05$ ). We here only represent the points that  are associated to
 positive $M[\bar{f}_{eq}]$. The transition occurs at constant energy $E_2 \simeq 0.675$, regardless of the specific
 domains that result in the two level water-bag distribution.}
\label{figure:b}
\end{figure}

To elucidate the specificity of the outlined transition, we plot in figure \ref{figure:b} the energy $E_2$, associated to each of the selected initial conditions, versus $f_1$, for the same selection of parameters as employed in figure \ref{figure:1}. As suggested by visual inspection of the figure, the  transitions, which we recall take place within a finite window in $f_1$, always occur for an identical value of the energy (in this case $E_2^c \simeq 0.675$). The transition point is hence insensitive to the specificity of the two water-bags, being neither dependent on their associated volumes nor relative heights. It is in principle possible to extend the above analysis 
and so reconstruct the complete transition surface in the ($f_1, f_2, E_2$) space, a task which proves however demanding from the computational viewpoint and falls outside the scope of the present paper.

\begin{figure}
\includegraphics[draft=false,clip=true, width=8cm]{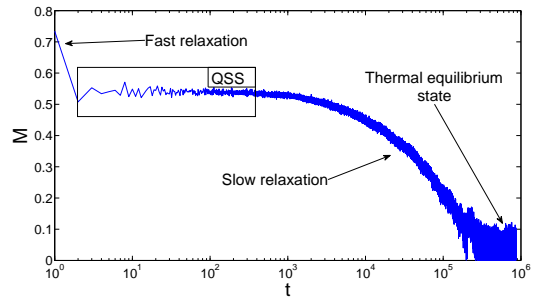}
\caption{Magnetization $M$ as a function of time t, as seen in a typical simulation. The system experiences a fast growth and then settle down into the lethargic QSS phase, whose duration (data not shown) increases with $N$. Later on the system moves towards the deputed equilibrium. In this simulation a two levels water-bag is assumed with
$f_1=0.14$, $f_2=0.1$, $\alpha_1= 0.2$, $\alpha_2= 0.3$. The energy is $E_2=1.0$ and $N=10^4$. }
\label{figureQSS}
\end{figure}

To test the validity of the theory we have run a series of numerical simulations of the HMF model. The implementation is based on fifth order McLachlan-Atela algorithm \cite{Atela} with a time-step $\delta t=0.1$. The initial condition is of a two levels water-bag type, with respective domains assigned as follows the aforementioned prescriptions.
 As a preliminary check we have monitored the approach to equilibrium, figure \ref{figureQSS}.

\begin{figure}
\includegraphics[draft=false,clip=true, width=8cm]{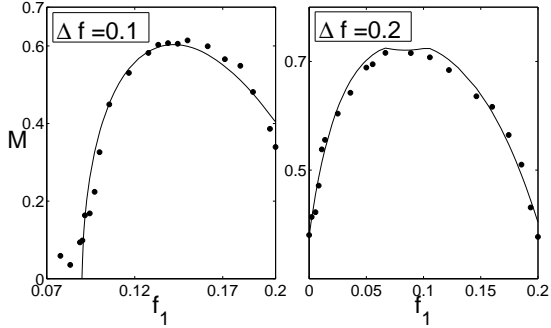}
\caption{The analytical predictions (solid line) for the QSS magnetization as a function of $f_1$ in the two-levels water-bag case, are compared (filled circles) to the numerical simulations performed for $N=10^4$. 
The comparison is drawn for two distinct values of $\Delta f$ ($\Delta f= 0.1$ (left) and  $\Delta f= 0.2$ (right)). $\alpha_1=5$ and $\alpha_2$ follows the normalization condition. Numerical values of $M$ are computed as a time average over a finite time window where the QSS holds.
The data are further mediated over $4$ independent realizations. Expected uncertainties are about the  
size of the circle.}
\label{figure:3}
\end{figure}

As expected and generalizing the conclusion that have been shown to hold for
the simpler one level water-bag family of initial conditions, the system
settles down into a QSS, whose lifetime grows with the number of simulated
particles (data not shown). 
 The QSS are indeed the target of our analysis and it is the magnetization as recorded in the QSS phase that needs to be compared to the Lynden-Bell predictions. The comparison between theory and simulations is reported in figure  
\ref{figure:3}. Filled symbols refer to the simulation while the solid line stand for the theory, for two distinct choice of $\Delta f$. The agreement is certainly satisfying and points to the validity of the Lynden-Bell interpretative framework, beyond the case of the single water-bag, so far discussed in the literature.

\section{Conclusions}

The dynamics of long range interacting system is studied, as concerns the intriguing emergence of long lasting Quasi Stationary States. The problem is tackled within the context of the Hamiltonian Mean Field model, a very popular and paradigmatic case study. Building on previous evidences, the QSS are interpreted as stable equilibria of the Vlasov equation, which rules the dynamics of the discrete HMF system in the infinite system size limit ($N \rightarrow \infty$). The QSS are hence characterized analytically by means of a maximum entropy principle inspired to the seminal work of Lynden Bell. This technique is known to yield to reliable predictions, when dealing with a very specific class of initial condition, the so called (single) water-bag. The scope of this paper is to push forward the analysis by considering the case where multiple water bags are allowed for. The theory is challenged with reference to the case of a two levels water-bag initial condition and the comparison with the simulations proves accurate. Phase transitions are in fact predicted and observed in direct $N$-body simulations. Motivated by 
this success, we argue that the Lynden-Bell approach could be adapted to more complex, and so realistic, family of  initial conditions.

\end{document}